\documentclass[11pt]{article}
\usepackage{latexsym}  
\usepackage{amssymb}
\usepackage{graphicx}
\usepackage{amsmath}

\begin{document}

\begin{center}

{\Large\bf Flavon VEV Scales in U(3)$\times$U(3)$'$ Model}

\vspace{4mm}

{\bf Yoshio Koide$^a$ and Hiroyuki Nishiura$^b$}

${}^a$ {\it Department of Physics, Osaka University, 
Toyonaka, Osaka 560-0043, Japan} \\
{\it E-mail address: koide@kuno-g.phys.sci.osaka-u.ac.jp}

${}^b$ {\it Faculty of Information Science and Technology, 
Osaka Institute of Technology, 
Hirakata, Osaka 573-0196, Japan}\\
{\it E-mail address: hiroyuki.nishiura@oit.ac.jp}

\date{\today}
\end{center}

\vspace{3mm}

\begin{abstract}
We have recently proposed a quark and lepton mass matrix model  
based on U(3)$\times$U(3)$'$ 
family symmetry as the so-called 
Yukawaon model, in which  the U(3) symmetry is broken 
by VEVs of flavons $(\Phi_f)_i^{\ \alpha}$ which are 
$({\bf 3}, {\bf 3}^*)$ of U(3)$\times$U(3)$'$. 
The model has successfully provided the unified 
description of quark and lepton masses and mixings
by using the observed charged lepton masses as only family-number
dependent input parameters. However, our final goal is not only to give 
well-satisfied fitting of quark and lepton masses and mixings, but 
to investigate physics behind such a successful parameter fitting.    
Therefore, our next concern is scales of VEVs of the flavons 
because we have not paid attention to scales of flavons in 
the previous study. In order to give consistency among the scales,
the previous flavon model is drastically changed together with economizing 
number of the flavons, but with keeping previous phenomenological success.   
We estimate that VEVs of flavons with $({\bf 8+1}, {\bf 1})$, 
$({\bf 3}, {\bf 3}^*)$, and $({\bf 1}, {\bf 8+1})$ are of 
the orders of $10$ TeV, $10^4$ TeV, and $10^7$ TeV, respectively.  
\end{abstract}

PCAC numbers:  
  11.30.Hv, 
  12.60.-i, 
  12.15.Ff, 
  14.60.Pq,  


\vspace{3mm}

{\large\bf 1  \  Introduction} 
 
The greatest concern in the flavor physics is how to understand
the origin of the observed hierarchical structures of  masses 
and mixings of quarks and leptons. 
Recently, we have proposed a quark and lepton mass matrix 
model based on U(3)$\times$U(3)$'$ flavor symmetry 
\cite{K-N_PRD15R, K-N_MPLA16}. 
The model is an extended version of the so-called ``Yukawaon" 
model \cite{Yukawaons}, which is a kind of flavon model \cite{flavon} 
to understand origin of masses 
(mass matrices) by vacuum expectation values (VEVs) of 
flavons.

In the past Yukawaon models, VEV scales of flavons have 
not been discussed. 
In this paper,  we take  the energy scales of VEV of flavons into consideration.
In order to discuss the VEV scales consistently, we have rebuilt the previous model. 
Especially, in the present model, ``Yukawaons" are absent in the limit of 
unbroken U(3)$\times$U(3)$'$ symmetry. 
We assume fermions $F_\alpha$ ($\alpha=1,2,3$) which are triplet of 
U(3)$'$ in addition to quark and leptons $f_i$ ($i=1,2,3$) which are
triplet of U(3), and we consider that  
the effective Yukawa coupling terms are caused by the $f$-$F$ mixing.   
As a result, the basic formulation of the model is drastically changed. 
However,  the successful phenomenological study of the masses and mixings
in the previous works \cite{K-N_PRD15R, K-N_MPLA16} is   
substantially kept
as far as quarks and leptons are concerned.  
Therefore, the purpose of the present paper  is not to 
discuss masses and mixings of quarks and leptons. 
Our concern is in the scales of U(3)$'$ and U(3) and 
a scale of the right-handed Majonara neutrino mass. 

Let us present the basic structure of the model:

(i) In the previous model with U(3)$\times$U(3)$'$ flavor symmetry, 
the quarks and leptons are assigned to
triplets of U(3) family symmetry. 

(ii)The Majorana neutrino mass matrix is considered based on the conventional 
seesaw mechanism \cite{seesaw}, 
$$
(M_\nu^{Maj})_{ij} = (\hat{M}_\nu)_i^{\ k} (M_R^{-1})_{kl} (\hat{M}_\nu)^l_{\ j} ,
\eqno(1.1)
$$
where $i,j=1,2,3$ are indexes of U(3) symmetry. Here 
the right-handed Majorana neutrino mass matrix $M_R$ will take 
a scale of U(3) symmetry breaking $\mu \sim \Lambda_3$. 

(iii) 
The difference among sectors 
$f=(u,d, \nu, e)$ seen in the observed masses and 
mixings is brought by scalars $S_f =({\bf 1}, {\bf 8}+{\bf 1})$ 
of U(3)$\times$U(3)$'$ by assuming that 
the U(3)$'$ symmetry is broken into a discrete 
symmetry S$_3$ at a scale $\mu=\Lambda_1$.  
The VEV forms of $S_f$ are given by 
$$
\langle \hat{S}_f \rangle = v_{S} ({\bf 1} + b_f X_3) ,
\eqno(1.2)
$$
where ${\bf 1}$ and $X_3$ are defined by
$$
{\bf 1} = \left( 
\begin{array}{ccc}
1 & 0 & 0 \\
0 & 1 & 0 \\
0 & 0 & 1 
\end{array} \right) , \ \ \ \ \ 
X_3 = \frac{1}{3} \left( 
\begin{array}{ccc}
1 & 1 & 1 \\
1 & 1 & 1 \\
1 & 1 & 1 
\end{array} \right) ,  
\eqno(1.3)
$$
and $b_f$ are complex parameters.  

(iv) The Dirac mass matrices $(\hat{M}_f)_i^{\ j}$ 
($f=u, d, \nu, e$) for quarks and leptons have been given by a common form 
$$
(\hat{M}_f)_i^{\ j}  = \langle\Phi_f\rangle_i^{\ \alpha}
\langle \hat{S}_f^{-1} \rangle_\alpha^{\ \beta} 
\langle \bar{\Phi}_f\rangle_\beta^{\ j} ,
\eqno(1.4)
$$
where $i, j$ and $\alpha, \beta$ are indexes of U(3) and
U(3)$'$, respectively.\footnote{ 
In this paper, we denote flavons with $({\bf 8}+{\bf 1}, {\bf 1})$ 
and $( {\bf 1}, {\bf 8}+{\bf 1})$ of U(3)$\times$U(3)$'$ 
as $\hat{A}_i^{\ j}$ and $\hat{A}_\alpha^{\ \beta}$ 
with ``hat", respectively.   
We also denote a flavon $({\bf 3}^*, {\bf 3}^*)$ as $\bar{A}^{\alpha i}$
in contrast to $A_{i \alpha}$ of  $({\bf 3}, {\bf 3})$.
However, for flavons with $({\bf 3}, {\bf 3}^*)$ and 
$({\bf 3}^*, {\bf 3})$, we denote those as $A_i^{\ \alpha}$ 
and $\bar{A}_\alpha^{\ i}$, respectively, giving priority 
to U(3) index.
}
In this model, all of scalars  (except for Higgs scalars $H$) 
are singlets of SU(3)$_c\times$SU(2)$ \times$U(1)$_Y$.  
Such a unified description of quark and lepton mass matrices 
based on the form (1.4) with matrix forms (1.2) and (1.3) 
has first been proposed in Ref.\cite{K-F_ZPC96}
in order to get top quark mass enhancement.  
(However, their work was not based on the symmetry  
U(3)$\times$U(3)$'$.)   
We consider that the form (1.4) comes from a seesaw-like 
mechanism for quarks and leptons $f_i$ and hypothetical 
heavy fermions $F_\alpha$. 
(The detail formulation is discussed in Sec.2.)

(v) The U(3) is  broken by 
VEVs of $\Phi_f$ with $({\bf 3}, {\bf 3}^*)$ of 
U(3)$\times$U(3)$'$, the VEV forms of  which are assumed to be diagonal as  
given by
$$
\langle \Phi_f \rangle = v_\Phi\, {\rm diag}( z_1 e^{i\phi^f_1}, 
z_2 e^{i\phi^f_2}, z_3 e^{i\phi^f_3}) .
\eqno(1.5)
$$
We shall see that $\phi_i^f=0$ for $f=d, e, \nu$ in later discussions. 
Therefore, we hereafter denote $(\phi_1^u, \phi_2^u, 
\phi_3^u)$ as $(\phi_1, \phi_2, \phi_3)$ simply.  
Namely, only for up-quark sector $f=u$, $\langle \Phi_u \rangle$ 
takes phase factors as $\langle \Phi_u \rangle = 
v_\Phi {\rm diag}(z_1 e^{i\phi_1} , z_2 e^{i\phi_2}, z_3 e^{i\phi_3})$. 

(vi) We will give $b_e=0$ for the parameter $b_e$ in Eq.(1.2) , i.e. $\hat{S}_e = v_S {\bf 1}$ as 
seen in Eq.(2.9) in the next section. 
Thereby the 
charged lepton mass matrix is given by 
$$
\hat{M}_e  \propto {\rm diag} (z_1^2, z_2^2, z_3^2) ,
\eqno(1.6)
$$
from which we get a relation
$$
z_i = \frac{ \sqrt{m_{ei}} }{\ \sqrt{ 
m_e + m_\mu + m_\tau} } .
\eqno(1.7)
$$
Here $m_{ei}=(m_e, m_\mu, m_\tau)$  are charged lepton masses, and 
the real parameters $z_i$ are normalized as 
$z_1^2+z_2^2+z_3^2=1$.  
Note that the parameter values $(z_1, z_2, z_3)$ are 
universal for every sector $f=u, d, e, \nu$. 
On the other hand, the parameters $b_f$ in Eq.(1.2) are 
sector-dependent ($f$-dependent) but family-independent. 
Family-dependent parameter $\phi_i$ are described in terms of $z_i$ 
\cite{K-N_PRD15} as seen in Sec.3.   
Therefore, family dependent parameters are 
only $z_i$. 
In this model, parameters which we can adjust are only 
the family-independent parameters $b_f$. 
Note that the parameters $b_f$ determine not 
only mass eigenvalues $(m_{f1}, m_{f2}, m_{f3})$ but also 
family mixing matrix $U_f$.
The origin of the parameter values (1.6) has not 
been given in the Yukawaon model.  
The relation (1.7) is nothing but an ad hoc assumption.

(vii) Finally let us  emphasize the strategy of the Yukawaon model 
based on the U(3)$\times$U(3)$'$ symmetry. 
In most mass matrix models, a symmetry for quarks and leptons is given, 
and thereby the masses and mixings are derived. 
(It is investigated how the family symmetry breaks into a 
suitable sub-symmetry or discrete symmetry.)
In the Yukawaon model, we do not propose an explicit symmetry 
breaking mechanism for U(3) symmetry. 
We only use the observed values of the charged lepton masses 
for the relation (1.7). 
The mechanism which gives charged lepton masses is left 
for a future task, and, for the moment, we do not ask 
the origin.
Instead, we try to give unified description of all mass ratios 
and mixings in terms of the charged lepton mass values, 
and without using any family-number dependent parameters at all. 
The purpose of the Yukawaon model is to show that all the quark and
lepton mass ratios and mixings can be described  when we accept   
the observed charged lepton mass values as only family-dependent
parameters. 

Anyhow, we could successfully obtain \cite{K-N_PRD15R} 
a unified description of 
quark and lepton masses and mixing by using the observed charged 
lepton masses as only family-number dependent input parameters.  
The aim has been almost accomplished in the phenomenological level. 
Therefore, the purpose of the present paper is not to give 
parameter fitting of quark and lepton masses and mixing, 
but to take  the energy scales of VEV of flavons into consideration 
consistently. 
In order to discuss the VEV scales consistently, 
the basic formulation of the model is drastically changed.

The new model in this paper is characterized as follows:
(i) There are no Yukawaons, although 
the flavons $\Phi_f$ and $\hat{S}_f$ still play an essential role
in the new model, too.  
(ii) There are three type of VEV scales.
We denote these scales as 
$$ 
\langle \hat{S}_f \rangle_\bullet^{\ \bullet} \sim \Lambda_1, \ \ \ \  
\langle\Phi_f\rangle_\circ^{\ \bullet} \sim \Lambda_2, \ \ \ \ 
\langle E_{\circ\circ}\rangle \sim \Lambda_3 .
\eqno(1.8)
$$
More exactly speaking, we define $\Lambda_1$ as $\Lambda_1 = v_{S}$ 
in Eq.(1.2), and $\Lambda_2$ as $\Lambda_2= v_\Phi$ (except for $f=\nu$) 
in Eq.(1.5), respectively.  
We have also denote a scale $\Lambda_3$ as $v_E =\Lambda_3$ 
which appears in Eq.(2.15) later.  
Here and hereafter, in order to be easy to see  a scale of flavons, 
we will sometimes show the indexes of U(3) and U(3)$'$,  by
``$\circ$" and ``$\bullet$", instead of $i, j, k, \cdots$" and 
"$\alpha, \beta, \gamma, \cdots$", respectively. 
We consider $\Lambda_1 \gg \Lambda_2 \gg \Lambda_3$. 
The purpose of the present paper is to fix those scales 
by the observed data. 
(iii) The most troublesome problem is VEV scales 
in neutrino sector. 
In oder to make the model anomaly free, the right-handed 
neutrinos $\nu_R$ belong to triplet of U(3) as well as $\nu_L$, 
so that the right-handed neutrino Majorana mass matrix 
$(M_R)^{ij}$ are given by VEV of a flavon $\bar{Y}_R$, 
$\langle \bar{Y}_R \rangle^{ij}$. 
We demand that Majorana mass matrix of the left-handed 
neutrinos is given by a seesaw mechanism 
$$
(M_{\nu}^{Maj})_{\circ\circ} = (\hat{M}_\nu)_\circ^{\ \circ} \langle 
\bar{Y}_R^{-1}\rangle_{\circ\circ}
(\hat{M}_\nu^T \rangle^\circ_{\ \circ} .
\eqno(1.9)
$$
However, 
a scale of the right-handed neutrino Majorana mass matrix  
is considerably small differently from the conventional 
neutrino seesaw \cite{seesaw}, 
because our flavon seems to have 
$\langle\bar{Y}_R\rangle^{\circ\circ} \sim \Lambda_3$. 
We will squarely grapple with this problem. 

In Sec.2, we discuss our Dirac mass matrix forms 
which are given in Eq.(1.4). 
In Sec.3, we discuss relations between phase parameters 
$\phi_i$ given in Eq.(1.5) and the charged lepton masses $m_{ei}$. 
That is, we show that the apparently ``family-number dependent 
parameters $\phi_i$ can be described in terms of charged lepton
mass parameters $(m_e, m_\mu, m_\tau)$.
This was first pointed out in Ref.\cite{K-N_PRD15}.  
In our present model, since the VEV relations are completely revised, 
the relations between $\phi_i$ and $m_{ei}$ are also 
completely different from previous ones.
The new relations lead us not only to that the relations give 
an understanding of the values $\phi_i$ on the basis 
of the charged lepton masses $m_{ei}$, but also to that
we can predict the scales $\Lambda_1$, $\Lambda_2$ and
$\Lambda_3$ by using the result of data-fitting in the 
$\phi_i$-$m_{ei}$ relations. 
In Sec.4, we discuss slight VEV deviations among flavons 
with the same transformation of U(3)$\times$U(3)$'$. 
This is not essential to predict $\Lambda_2/\Lambda_1$ 
and $\Lambda_3/\Lambda_2$. 
However, in order to obtain more consistency in our scenario,
we will discuss somewhat detailed VEV relations. 
In Sec.5, we discuss a Mojorana mass matrix form of the 
right-handed neutrinos $\nu_R$ as a preparation for  
predicting the scales $\Lambda_1$, $\Lambda_2$ and
$\Lambda_3$. 
In Sec.6, we discuss the VEV scales $\Lambda_1$, $\Lambda_2$ 
and $\Lambda_3$. 
Finally, Sec.7 is devoted to concluding remarks. 
In Appendix A, the detail papameter fitting for  the masses and mixings 
of quarks and leptons presented in Ref \cite{K-N_PRD15R, K-N_MPLA16} is reviewed. 

\vspace{5mm}

{\large\bf 2 \ Dirac mass matrices} 

\vspace{2mm}

{\bf 2.1 \ Seesaw-type mechanism in the Dirac mass matrix}

In our model based on U(3)$\times$U(3)$'$ symmetry, 
we consider hypothetical fermions $F_\alpha$ 
($\alpha=1,2,3$), which belong to $({\bf 1}, {\bf 3})$ of
 U(3)$\times$U(3)$'$, 
in addition to quarks and leptons $f_i$ ($i=1,2,3$) 
which belong to  $({\bf 3}, {\bf 1})$.   
In this model, differently from the Yukawaon formulation, 
we assume that the VEV form (1.4) 
originates from the following $6\times 6$ mass matrix model:
$$
(\bar{f}_L^i \ \ \bar{F}_L^\alpha ) 
\left(
\begin{array}{cc}
(0)_i^{\ j}  &  (\Phi_f)_i^{\ \beta}  \\
(\bar{\Phi}_{f})_\alpha^{\ j} & -(S_f)_\alpha^{\ \beta} 
\end{array} \right) 
 \left(
\begin{array}{c}
f_{Rj} \\
F_{R\beta}
\end{array} 
\right) .
\eqno(2.1)
$$
Here, $F_{L(R)}$ are heavy fermions with 
$({\bf 1}, {\bf 1}, {\bf 3})$ of
SU(2)$_L \times$U(3)$\times$U(3)$'$.
On the other hand, 
$f_R$ are right-handed quarks and leptons,
$f_R= (u, d, \nu, e^-)_R$, while
$f_L$ are not physical fields.  
They are given by the following combinations: 
$$
f_L \equiv (f_u, f_d, f_\nu, f_e)_L \equiv 
\left( \frac{1}{\Lambda_H} H_u^\dagger q_L, 
\frac{1}{\Lambda_H} H_d^\dagger q_L,
\frac{1}{\Lambda_H} H_u^\dagger \ell_L,
\frac{1}{\Lambda_H} H_d^\dagger \ell_L \right)
\eqno(2.2)
$$
where 
$$
q_L = \left(
\begin{array}{c}
u_L \\
d_L 
\end{array} \right) , \ \ \ \ 
\ell_L = \left(
\begin{array}{c}
\nu_L \\
e^-_L 
\end{array} \right) , \ \ \ \ 
H_u = \left(
\begin{array}{c}
H_u^0 \\
H_u^- 
\end{array} \right) , \ \ \ \ 
H_d = \left(
\begin{array}{c}
H_d^+ \\
H_d^0 
\end{array} \right) . 
\eqno(2.3)
$$
In other words, the matrix given in Eq.(2.1) denotes would-be 
Yukawa coupling constants. 

After the U(3) and U(3)$'$ have been completely broken, 
the quarks and leptons are described by the effective 
Hamiltonian
$$
{\cal H}_Y = (\bar{\nu}_L)^i (\hat{M}_\nu)_i^{\ j}  (\nu_R)_j 
+ (\bar{e}_L^i (\hat{M}_e)_i^{\ j} (e_R)_j  
+ y_R (\bar{\nu}_R)^i (Y_R)_{ij} (\nu_R^c)^j 
$$
$$
+  (\bar{u}_L)^i (\hat{M}_u)_i^{\ j} (u_R)_j 
+ (\bar{d}_L)^i (\hat{M}_d)_i^{\ j}  (d_R)_j  .
\eqno(2.4)
$$
Of course, $(\hat{M}_f)_i^{\ j}$ do not mean flavons although 
we have called those ``Yukawaons" in the past Yukawaon model
\cite{Yukawaons}. 
Note that the quarks and leptons $f_i$ are not U(3) family 
triplet any more in the exact meaning, but they are mixing states between 
$f$ and $F$.
However, we will still use the index of U(3) family for these 
fermion states. 
Also, note that there are no Yukawaons in the present model. 
By performing a seesaw-like approximation with $\Lambda_2 \ll \Lambda_1$, 
the mass matrix (2.1) leads to the following Dirac mass matrices 
of quarks and leptons:
$$
(\hat{M}_f)_i^{\ j} \simeq \frac{\langle H_{u/d} \rangle}{\Lambda_H} 
 \langle {\Phi}_{f} \rangle_i^{\ \alpha} 
\langle (\hat{S}_f)^{-1} \rangle_{\alpha}^{\ \beta} 
\langle \bar{\Phi}_{f} \rangle_{\beta}^{\ j} .
\eqno(2.5)
$$

In this paper, we search VEV scales of flavons by assuming 
that our flavon VEV scale is given by one of $\Lambda_1$, 
$\Lambda_2$ and $\Lambda_3$, except for $H_{u/d}$. 
As we stated in subsection.2.3, we put $1/(\Lambda_1)^n$ for 
superpotential term with dimension $(3+n)$.
However, we assume that these rules are exceptional 
for the factor $H_{u/d}/\Lambda_H$. 
The VEV of $H_u$ and $H_d$ are fixed as 
$$
v_{Hu} = v_{Hd} \equiv v_H \equiv \frac{1}{\sqrt{2}} 
\times 246 \ {\rm GeV}.
\eqno(2.6)
$$
(The reason of $v_{Hu} = v_{Hd}$ will be given in 
Eq.(4.6) later.)
We will take $\Lambda_H$ as a different value from 
$\Lambda_1$, $\Lambda_2$ and $\Lambda_3$. 
That is, we consider that the additional factor 
$H_{u/d}/\Lambda_H$ in Eq.(2.5) is  factor at the 
electroweak scale after our family symmetries 
U(3)$\times$U(3)$'$ are completely broken. 
For the time being, we do not discuss the mechanism 
which gives the factor $H_{u/d}/\Lambda_H$. 

\vspace{2mm}

{\bf 2.2 \ $R$-charge assignment of $\Phi_f$} 

In this section, we will often mention an effective superpotential.
However, differently from the conventional SUSY scenario, our 
SUSY is already broken at a high energy scale, for example, 
$\mu \sim 10^7$ TeV. 
Nevertheless, we assume that interaction forms at the unbroken 
SUSY scale are still kept in low energy (smaller than $\mu < 10^4$ 
TeV) phenomenology.   

We assume that $R$ charges are still conserved under the block 
diagonalization of (2.1), so that we consider an $R$ charge relation
$$
R(\hat{S}_f) = R(\Phi_f) \equiv r_f .
\eqno(2.7)
$$
Here we denote $R$ charge of the flavon $\hat{S}_f$ as $R(\hat{S}_f)$ and so on.
We have taken $R$ charge of the Higgs fields as 
$R(H_{u/d})=0$. 
Then, the Dirac mass matrix $\hat{M}_f$ 
has effectively 
a $R$ charge $r_f$ (although $\hat{M}_f$ is not a flavon).
Note that if there is a flavon combination 
$(\Phi_{f'}\bar{\Phi}_{f^{\prime\prime}})_i^{\ j}/\Lambda_1$
with $r_{f'}+r_{f^{\prime\prime}} = r_f$, 
the combination can couple to $\bar{f}_L^j f_{Ri}$ 
as well as $(\hat{M}_f)_i^{\ j}$ with the same 
energy scale $(\Lambda_2)^2/\Lambda_1$. 
Besides, if $(\bar{\Phi}_f \Phi_{f'})$ and another 
$(\bar{\Phi}_{f"} \Phi_{f"'})$ have the same $R$ 
charge, those can have the same VEV structure. 
This is an unwelcome situation, because we demand that 
$\Phi_u$ and $\Phi_\nu$ have somewhat different VEV 
structures from $\Phi_e$ and $\Phi_d$ as we state later.
A simple way to avoid appearance of such unwelcome 
combinations is to take $R$ charge difference among 
$\Phi_f$'s completely different from each other:
$r_e-r_u=2/3$, $r_d-r_e = 3/3$, $r_\nu -r_d =4/3$, 
i.e. 
$$
(r_u, r_e, r_d, r_\nu) = \left( \frac{1}{3}, 1, 2,
\frac{10}{3} \right) .
\eqno(2.8)
$$
The reason $r_u < r_e$ will also be discussed later.

The reason why we choose $r_e =1$ is as follows:
Only when $r_e=1$, we can write down a superpotential for 
$\hat{S}_e$,
$$
W_{Se} = \lambda^S_{1}[(\hat{S}_e)_\alpha^{\ \beta} 
(\hat{S}_e)_\beta^{\ \alpha} ] + 
 \lambda^S_{2}[(\hat{S}_e)_\alpha^{\ \alpha} ]
[(\hat{S}_e)_\beta^{\ \beta} ] . 
\eqno(2.9)
$$
Supersymmetric vacuum condition for this $W_{Se}$ leads to $\hat{S}_e = v_S {\bf 1}$,
so that we obtain $b_e = 0$. 
This means that the charged lepton mass matrix $\hat{M}_e$ is 
proportional to ${\rm diag}(z_1^2, z_2^2, z_3^2)$.
Therefore, our parameters $z_i$ are given by (1.7).
The phase parameters $\phi_i^e$ do not have physical meaning 
because $\langle\Phi_e\rangle$ is commutable with 
$\langle \hat{S}_e\rangle = v_S {\bf 1}$, so that we can 
hereafter put $\phi_i^e =0$, i.e. 
$\langle \Phi_e \rangle = v_e\, {\rm diag}(z_1, z_2, z_3)$. 

In Table.1 we list $R$ charges of leading flavons. 

\vspace{2mm}

{\bf 2.3 \ VEV relation among $\Phi_f$}

 In this paper, we will often give effective superpotential terms. 
We write down those effective superpotentials according to the 
following rules:
(i) For effective superpotential terms with a higher mass-dimension 
$(n+3)$, we put a factor $(1/\Lambda_1)^n$ in order to adjust 
the term to the dimension 3.   
(ii) In order to avoid unwelcome terms with higher dimension, 
we assume $R$ charge conservation and positive $R$ charge 
assignment (except for special flavons $\Theta$ as we show later),
so that superpotential terms with $R >2$ are forbidden. 

A VEV of $\Phi_d$ is obtained by the following superpotential
$$
W_d = \left\{ (\Phi_d)_i^{\ \beta} (\hat{E})_\beta^{\ \alpha} +
\lambda_d \frac{1}{\Lambda_1} (\Phi_e)_i^{\ \beta} 
(\hat{S}_e)_\beta^{\ \gamma} (\hat{E})_\gamma^{\ \alpha} 
\right\} (\bar{\Theta}_d)_\alpha^{\ i} .
\eqno(2.10)
$$
Here and hereafter, we assume that the $\Theta$ fields always 
take $\langle \Theta\rangle =0$.  
Therefore, from $\partial W_d/\partial \Theta_d=0$, 
we obtain 
$$
(\Phi_d)_i^{\ \beta} (\hat{E})_\beta^{\ \alpha} =
-\lambda_d \frac{1}{\Lambda_1} (\Phi_e)_i^{\ \beta} 
(\hat{S}_e)_\beta^{\ \gamma} (\hat{E})_\gamma^{\ \alpha} .
\eqno(2.11)
$$
Therefore, when the scale of $(\hat{E})_\gamma^{\ \alpha}
$ is $\Lambda_1$, 
we obtain the VEV $\langle \Phi_d \rangle$ with
the same order as $\langle \Phi_e \rangle$, i.e. 
$\langle \Phi_d\rangle = v_d\,{\rm diag}(z_1, z_2, z_3)$ 
with $\phi_i^d=0$. 

However, for $\Phi_u$ with $R=1/3$, we have to consider 
another type of  superpotential,
$$ 
W_u = \left\{ \frac{\lambda_{u1}}{\Lambda_1} (\bar{\Phi}_u)_\alpha^{\ k}
(\Phi_u)_k^{\ \gamma} (\hat{E})_\gamma^{\ \beta} +
\lambda_{u2}(\bar{\Phi}_d)_\alpha^{\ k}(\Phi_d)_k^{\ \beta} 
\right\} (\hat{\Theta}_u)_\beta^{\ \alpha} . 
\eqno(2.12)
$$
Here, we have introduced a new flavon $\hat{E}_\bullet^{\ \bullet}$
with $R=5/3$ and a VEV form $\langle \hat{E} \rangle =v_E {\bf 1}$. 
Since this potential gives a relation between $\Phi_u \bar{\Phi}_u$
and  $\Phi_d \bar{\Phi}_d$, $\langle \Phi_u \rangle$ can have 
phase factors differently from real VEV matrix $\langle \Phi_d \rangle$.
Hereafter, we simply denote phase parameters $\phi_i^u$ as $\phi_i$:
$$
\langle \Phi_u \rangle = v_u\, {\rm diag}( z_1 e^{i\phi_1}, 
z_2 e^{i\phi_2}, z_3 e^{i\phi_3}) .
\eqno(2.13)
$$ 

Meanwhile, in this model, flavons $E$ with VEV matrix form
$v_E {\rm diag}(1, 1, 1)$ frequently appear.  
For $E$ with the VEV scale $\Lambda_3$ and $R=2/3$, we consider a superpotential given by
$$
W_E = \lambda_1 {\rm Tr}[\hat{E}_\circ^{\ \circ} E_{\circ\circ}
\bar{E}^{\circ\circ}] + \lambda_2 {\rm Tr}\hat{[E}_\circ^{\ \circ}]
{\rm Tr}[ E_{\circ\circ}\bar{E}^{\circ\circ}] .
\eqno(2.14)
$$
SUSY vacuum condition leads to 
$$
\hat{E}_\circ^{\ \circ} = v_E {\bf 1}, \ \ \ \ 
  E_{\circ\circ}\bar{E}^{\circ\circ} = (v_E)^2 
  {\bf 1} .
\eqno(2.15)
$$
For $E_{\circ\circ}$, we take a special solution 
 $E_{\circ\circ} = v_E {\bf 1}$ in the relation
$E_{\circ\circ}\bar{E}^{\circ\circ} = 
(v_E)^2  {\bf 1}$. 
For $\hat{E}_\bullet^{\ \bullet}$ and $E_{\bullet\bullet}$,  
there is not such a simple superpotential. 
We have to assume the following superpotential
$$
W_E = \lambda_1 {\rm Tr}[\hat{E}_\bullet^{\ \bullet}
\hat{E}_\bullet^{\ \bullet} 
(\hat{\Theta}_{-4/3})_\bullet^{\ \bullet}]
+ \lambda_2 {\rm Tr}[ E_{\bullet\bullet} \bar{E}^{\bullet\bullet}
(\hat{\Theta}_{-4/3})_\bullet^{\ \bullet}] + \dots,
\eqno(2.16)
$$
where $(\hat{\Theta}_{-4/3})_\bullet^{\ \bullet} $ has $R=-4/3$. 
We also assume 
$$
W_E = {\rm Tr}[\hat{E}_\circ^{\ \circ} E_{\circ\bullet} 
\bar{E}^{\bullet\circ} (\hat{\Theta}_{2/3})_\circ^{\ \circ}] 
+\cdots, 
\eqno(2.17)
$$ 
where $(\hat{\Theta}_{2/3})_\circ^{\ \circ} $ has $R=2/3$.

\begin{table}
\caption{ 
$R$ charges and VEV scales of leading flavons. 
Transformation property under U(3)$\times$U(3)$'$ is 
indicated by ``$\circ$" and ``$\bullet$", respectively. 
}
\begin{center}
\begin{tabular}{|c|cccc|cccc|}\hline
flavon   & $(\Phi_u)_\circ^{\ \bullet}$ & $(\Phi_e)_\circ^{\ \bullet}$ &
$(\Phi_d)_\circ^{\ \bullet}$ & $(\Phi_\nu)_\circ^{\ \bullet}$ &
$(\hat{S}_u)_\bullet^{\ \bullet}$ & $(\hat{S}_e)_\bullet^{\ \bullet}$ & 
$(\hat{S}_d)_\bullet^{\ \bullet}$ & $(\hat{S}_\nu)_\bullet^{\ \bullet}$
\\ \hline 
$R$ & $\frac{1}{3}$ & $\frac{3}{3}$ & $\frac{6}{3}$ & 
$\frac{10}{3}$ & $\frac{1}{3}$ & $\frac{3}{3}$ & $\frac{6}{3}$ & 
$\frac{10}{3}$ 
\\[0.01in] 
Scale & $\Lambda_2$ & $\Lambda_2$ & $\Lambda_2$ & $\Lambda_3$ &
 $\Lambda_1$ & $\Lambda_1$ & $\Lambda_1$ & $\Lambda_2$ 
\\ \hline
\end{tabular}

\begin{tabular}{|cccccccc|}\hline
$E_{\circ\circ}$ & $\hat{E}_\circ^{\ \circ}$ & $E_{\circ\bullet}$ & 
$E_{\bullet\bullet}$ & $\hat{E}_\bullet^{\ \bullet}$ &
$(\hat{Y}_{eu})_\bullet^{\ \bullet}$ & $(\bar{S}'_u)^{\circ\bullet}$ 
& $(\bar{Y}_R)^{\circ\circ}$
\\ \hline
$\frac{2}{3}$   & $\frac{2}{3}$ & $\frac{1}{3}$ & $\frac{5}{3}$ & 
$\frac{5}{3}$ & $\frac{4}{3}$ & $\frac{5}{3}$ & $\frac{13}{3}$
\\[0.01in] 
$\Lambda_3$ & $\Lambda_3$ & $\Lambda_2$ & $\Lambda_1$ & $\Lambda_1$ 
& $\Lambda_1$ & $\Lambda_2$ & $\Lambda_3$
\\[0.01in]\hline
\end{tabular}

\begin{tabular}{|cccccccccc|}\hline
$(\bar{\Theta}_d)_\bullet^{\ \circ}$ & $\bar{\Theta}_u)_\bullet^{\ \bullet}$
 & $(\bar{\Theta}_\nu)_\bullet^{\ \circ}$ & 
 $(\hat{\Theta}_\nu)_\bullet^{\ \bullet}$ & 
 $(\hat{\Theta}_\phi)_\bullet^{\ \bullet}$ &
$({\Theta}_R)_{\circ\circ}$ & 
$(\hat{\Theta}_{eu})_\bullet^{\ \bullet}$ & 
$({\Theta}'_{Su})_{\bullet\circ}$ 
& $(\hat{\Theta}_E)_\bullet^{\ \bullet}$ & 
$(\hat{\Theta}_{E})_\circ^{\ \circ}$ 
\\ \hline
$-\frac{5}{3}$   & $-\frac{2}{3}$ & $-\frac{4}{3}$ & $-\frac{4}{3}$ & 
$\frac{2}{3}$ & $ -3$ & $\frac{2}{3}$ & $-1$ & 
$-\frac{4}{3}$ & $\frac{2}{3}$
\\[0.01in] 
$0$ & $0$ & $0$ & $0$ & $0$ & $0$ & $0$ & $0$ & $0$ & $0$ \\ \hline
\end{tabular}

\end{center}
\end{table}


Finally, let us discuss a VEV relation of $\langle \Phi_\nu \rangle$. 
We cannot consider the similar mechanism as   
$\langle \Phi_f \rangle$ using $\hat{E}$,
since we need to give a small scale compared with 
$\langle \Phi_e \rangle \sim \Lambda_2$ 
in order to satisfy the seesaw approximation (1.9). 
(Since the scale $\Lambda_2$ means the maximal scale 
of a flavon with $A_\circ^{\ \bullet}$, the requirement
$\langle \Phi_\nu \rangle < \Lambda_2$ has no problem.)   
Therefore, we assume the following superpotential term:
$$
W_\nu = \left\{ \mu_\nu (\Phi_\nu)_\circ^{\ \bullet} +
\lambda_\nu \frac{1}{(\Lambda_1)^2 } ({\Phi}_e)_\circ^{\ \bullet} 
\bar{E}^{\bullet \bullet} E_{\bullet\circ} 
\bar{E}^{\circ\bullet} 
\right\} (\bar{\Theta}_\nu)_\bullet^{\ \circ} ,
\eqno(2.18)
$$
where new flavons $E_{\circ\circ}$ and 
 $E_{\bullet\bullet}$ with VEV form $v_E {\bf 1}$ has $R$ charges
$$
R(E_{\circ\circ})=  \frac{2}{3}, \ \ \ \ 
R({E}_{\bullet\bullet}) = \frac{5}{3}, \ \ \ \
R({E}_{\circ\bullet}) = \frac{1}{3}. 
\eqno(2.19)
$$
Therefore, a VEV scale of $(\Phi_\nu)_\circ^{\ \bullet}$ with $R=10/3$
is given by 
$$
\| \Phi_\nu \| = \frac{(\Lambda_2)^3}{\mu_\nu  \Lambda_1} ,  
\eqno(2.20)
$$
differently from other $(\Phi_f)_\circ^{\ \bullet}$. 
Hereafter, for convenience, we denote a VEV scale of 
a flavon $A$ as a notation $\|A\|$. 
Also, in the estimation of VEV scales, for simplicity, 
we put $\lambda = 1$ for all 
dimensionless coefficients $\lambda$ in superpotentials 
(2.9), (2.10), and so on.
We denote Eq.(2.20) as
$$
 \langle \Phi_\nu \rangle = \xi_\nu \langle \Phi_e \rangle, 
 \ \ \ \ \xi_\nu = \frac{(\Lambda_2)^2}{\mu_\nu \Lambda_1} .
 \eqno(2.21)
$$ 

We have obtained $b_e=0$, i.e. $\langle \hat{S}_e \rangle 
= v_S {\bf 1}$ by assigning $r_e=1$ as shown in Eq.(2.9). 
However, since $\hat{S}_\nu$ has $R=10/3$, we cannot 
assert $b_\nu=0$ by means of a similar way to Eq.(2.9). 
We want a similar VEV structure of $\hat{M}_\nu$ to 
$\hat{M}_e$ except for its VEV scale. 
Therefore, we assume a superpotential similar to (2.18):
$$
W_{S\nu} = \left\{ \mu_{S\nu} (\hat{S}_\nu)_\bullet^{\ \bullet} +
\lambda_\nu \frac{1}{(\Lambda_1)^2 } (\hat{S}_e)_\bullet^{\ \bullet} 
\bar{E}^{\bullet \bullet} E_{\bullet\circ} \bar{E}^{\circ \bullet} 
\right\} (\hat{\Theta}_\nu)_\bullet^{\ \bullet} .
\eqno(2.22)
$$ 
Then, we obtain a result similar to (2.21):
$$
 \langle \hat{S}_\nu \rangle = \xi_{S\nu} \langle \hat{S}_e \rangle, 
 \ \ \ \ \xi_{S\nu} = \frac{(\Lambda_2)^2}{\mu_{S\nu} \Lambda_1} ,
 \eqno(2.23)
$$
so that we can obtain  $\hat{M}_\nu$ with the same form as
$\hat{M}_e$:
$$
  \hat{M}_\nu  = \xi_{M} \hat{M}_e , 
 \ \ \ \ \xi_{M} = (\xi_\nu)^2 (\xi_{S\nu})^{-1} =
 \frac{\mu_{S\nu} (\Lambda_2)^2}{(\mu_{\nu})^2 \Lambda_1} .
 \eqno(2.24)
$$

\vspace{2mm}

{\bf 2.4 \ Parameter values in the quark sector} 

As far as quark sector is concerned, the VEV matrices are 
exactly the same as those in the previous paper, 
although the model is completely different from the 
previous one.

We choose parameter values\footnote{
In the parameter fitting, we have used running masses at 
$\mu= m_Z$ as input quark and lepton masses.
}
as
$$
b_u = -1.011 , \ \ \ \ b_d = -3.522 \, e^{i\ 17.7^\circ},            
\eqno(2.25)
$$
and
$$
(\tilde{\phi}_1, \tilde{\phi}_2 )= ( -176.05^\circ,\ -167.91^\circ) , 
\eqno(2.26)
$$
where we have fitted $(\phi_1, \phi_2, \phi_3)$ as 
$(\tilde{\phi}_1, \tilde{\phi}_2, \ 0 )$ without losing generality.  
Then, we can obtain reasonable quark masses and 
Cabibbo-Kobayashi-Maskawa (CKM) mixing \cite{CKM} 
as shown in the previous papers \cite{K-N_PRD15R, K-N_MPLA16} as shown in Appendix A.  
Note that the parameter $\phi_0$ cannot be fixed in the 
CKM fitting, i.e, it is an unobservable parameter as we 
discuss in the next section. 

\vspace{5mm}

{\large\bf 3 \ Relation between phase parameters $\phi_i$ 
and charged lepton masses $m_{ei}$} 

The parameters $(\phi_1, \phi_2, \phi_3)$ were typical 
family-number dependent parameters. 
Now we try to describe $(\phi_1, \phi_2, \phi_3)$ in terms
of charged lepton masses $m_{ei}$ and two family-number 
independent parameters. 
This has first been pointed out by the authors \cite{K-N_PRD15}.
However, since the $R$-charge assignment in the present model 
is completely different from that in the previous model, 
the relations between the phase parameters $\phi_i$ and 
charged lepton masses $m_{ei}$ is also changed. 
New relations will be simply denoted with fractional 
coefficients and thereby it will make possible to 
estimate the scales of U(3)$\times$U(3)$'$.   
 
We assume the following superpotential
$$
W_{\phi} = \left\{ \frac{\lambda_1^{\phi}}{\Lambda_1} 
\left[ (\bar{\Phi}_u)_\bullet^{\ \circ} E_{\circ\circ} 
(\bar{E})^{\circ\bullet} + h.c. \right] +
\frac{\lambda_{2}^{\phi}}{(\Lambda_1)^2} 
(\bar{\Phi}_u)_\bullet^{\ \circ} ({E})_{\circ\bullet} 
(\bar{E})^{\bullet\circ} (\Phi_u)_\circ^{\ \bullet}
 \right.
$$
$$
\left.
+ \frac{\lambda_{2}^{\phi}}{(\Lambda_1)^3} 
 (\bar{\Phi}_u)_\bullet^{\ \circ} (\Phi_u)_\circ^{\ \bullet} 
  (\bar{\Phi}_u)_\bullet^{\ \circ} (\Phi_u)_\circ^{\ \bullet} 
\right\} (\hat{\Theta}_\phi)_\bullet^{\ \bullet} ,
\eqno(3.1)
$$
where $R(\Theta_\phi)=2/3$.  
Note that in addition to the second term $(\bar{\Phi}_u)_\bullet^{\ \circ} 
 ({E})_{\circ\bullet} (\bar{E})^{\bullet\circ} 
(\Phi_u)_\circ^{\ \bullet}$ in Eq.(3.1),   
the following additional terms are 
allowed: 
$(\bar{\Phi}_u)_\bullet^{\ \circ} (\Phi_u)_\circ^{\ \bullet}
 ({E})_{\bullet\circ} (\bar{E})^{\circ\bullet}$,
$(\bar{\Phi}_u)_\bullet^{\ \circ} 
 ({E})_{\circ\bullet} (\Phi_u^T)^\bullet_{\ \circ}
(\bar{E})^{\circ\bullet}$, 
$ ({E})_{\bullet\circ}  (\bar{\Phi}_u^T)^\circ_{\ \bullet} 
\bar{E}^{\bullet\circ} (\Phi_u)_\circ^{\ \bullet}$,
and $ ({E})_{\bullet\circ}   (\bar{E})^{\circ\bullet}
(\bar{\Phi}_u)_\bullet^{\ \circ} (\Phi_u)_\circ^{\ \bullet}$.
Here, some remarks are in order:  (i) we regard those additional 
terms as ``substantially same terms",  
(ii) but, we count a flavon $A^\dagger$ as a different
field from $A$, and    
(iii) the coefficient $\lambda$ is defined as follows:
the $\lambda$ is a coefficient with 
a factor $1/n$ for sum of $n$ substantially same terms.
For example, in the second term in (3.1), the factor 
$\lambda_2^\phi$ is defined as one for sum of the 
five terms with $1/5$. 
However, for simplicity, we denote only 
representative one even if there are many equivalent terms,
and give the coefficient $\lambda$ instead of $\lambda/n$. 
Also note that the $h.c.$ term in the first term is 
different from the original one according to our counting rule.  

The flavon $\hat{\Theta}_\phi$ in (3.1) has 
VEV value of zero.
The SUSY vacuum condition $\partial W_\phi/\partial \Theta_\phi=0$
leads to a condition
$$
2 c_1 z_i \cos \phi_i = c_2 z_i^2 + c_3 z_i^4 , 
\eqno(3.2)
$$
where parameters $c_1$, $c_2$ and $c_3$ are family-number independent 
parameters and they have scales
$$
c_1 = \lambda_1 \frac{(\Lambda_2)^2 }{(\Lambda_1)^2}\Lambda_1 \Lambda_3, 
\ \ \ \  
c_2 = \lambda_2 \frac{(\Lambda_2)^4 }{(\Lambda_1)^2}, \ \ \ \ 
c_3 = \lambda_3 \frac{(\Lambda_2)^4 }{(\Lambda_1)^2}.
\eqno(3.3)
$$

When we denote these parameters $\phi_i$ as
$$
\begin{array}{l}
\phi_1 = \phi_0 + \tilde{\phi}_1 ,  \\
\phi_2 = \phi_0 + \tilde{\phi}_2 ,  \\
\phi_3 = \phi_0  ,  
\end{array}
\eqno(3.4)
$$
the parameter $\phi_0$ is unobservable in the CKM parameter 
fitting, while it is not unobservable in the 
U(3)$\times$U(3)$'$ model. 
From the input values 
$(\tilde{\phi}_1, \, \tilde{\phi}_2)= (-176.05^\circ,\ -167.91^\circ)$ given in (2.26),
we obtain
$$
\phi_0 = 86.69^\circ, \ \ \ \frac{c_2}{c_1}= 1.368, \ \ \ 
\frac{c_3}{c_2}=-0.967 ,
\eqno(3.5)
$$
which leads to
$$
(\phi_1,\, \phi_2,\, \phi_3) =(-89.36^\circ,\, -87.25^\circ,\, 
86.69^\circ) .
\eqno(3.6)
$$
Thus, the family-number dependent parameters 
$(\phi_1, \phi_2, \phi_3)$ can be reduced into 
family-number independent parameters $(c_2/c_1, c_3/c_2)$.  
(Note that the parameter $\phi_0$ is not unobservable 
 any longer in this model.) 

Note that the numerical results (3.5) suggests $|c_3/c_2| = 1$
if we take $\lambda_2=\lambda_3$.
Then, it is natural that we consider 
$\lambda_1=\lambda_2 =\lambda_3$.
Only the ratio $c_2/c_1$ is not one.
When we compare (3.3) with the numerical result  $c_2/c_1 \equiv \rho \simeq 1.37$ given in (3.5), we obtain
$$
R_\Lambda \equiv \frac{\Lambda_2}{\Lambda_1} = 
\rho  \frac{\Lambda_3}{\Lambda_2} .
\eqno(3.7)
$$ 
We will see $R_\Lambda \sim 10^{-3}$ in Sec.5. 



\vspace{5mm}

{\large\bf 4 \ Slight deviation of the scales in the lepton sector}

\vspace{2mm}

{\bf 4.1  $\hat{M}_{lepton}$ versus  $\hat{M}_{quark}$}

Let us comment on the ratio 
$v_{Hu}/v_{Hd}$.
Usually, it is understood that the value of 
$\tan \beta\equiv v_{Hu}/v_{Hd}$ is $\tan\beta \sim 10 - 40$. 
However, from (2.5), our quark masses are predicted as
$$
\begin{array}{l}
(m_u, m_c, m_t) = (0.0003964,\ 0.106411,\ 29.743) m_{0u} , \\
(m_d, m_s, m_b) = (0.0007249,\ 0.01467,\ 0.5365) m_{0d} ,
\end{array}
\eqno(4.1)
$$
where $m_{0f}$ are defined by 
$$
m_{0f} =\frac{v_{Hf}}{\Lambda_H} \frac{(v_f)^2}{v_{Sf}} 
= \frac{v_{Hf}}{\Lambda_H} \rho \Lambda_3 .
\eqno(4.2)
$$
In the last term in Eq.(4.2),  we have used the relation 
(3.7).  
The numerical values in (4.1) are eigenvalues 
of the dimensionless matrix 
$$
Z ({\bf 1} + b_f X_3)^{-1} Z,
\eqno(4.3)
$$
where 
$$
Z= {\rm diag}(z_1, z_2, z_3), 
\eqno(4.4)
$$
and we have used input values $b_u = -1.011$ and 
$b_d= -3.3522 e^{i\, 17.7^\circ}$ \cite{K-N_PRD15R}. 

The prediction (4.1) can give reasonable values not only 
for mass ratios $m_u/m_c$, $m_c/m_t$,  $m_d/m_s$, and 
$m_s/m_b$, but also for the mass ratio of up-quark mass to 
down-quark mass
$$
\frac{m_t}{m_b} = 55.44 ,  
\eqno(4.5)
$$
if we take $m_{0u}=m_{0d}$. 
The value (4.5) roughly agrees with the observed 
value \cite{q-mass} $({m_t}/{m_b})^{obs} \simeq 59.41$.
Therefore, in the present model, we can regard 
$$
\frac{m_{0u}}{m_{0d}}=1, \ \ \ {i.e} \ \ \  
v_{Hu} = v_{Hd} .
\eqno(4.6)
$$
Therefore, we have already taken $v_{Hu} = v_{Hd}$ in 
Eq.(2.6). 

Also, since $m_\tau = (z_3)^2 m_{0e}$ 
($z_3=0.97170$), we obtain
$$
m_{0e} = 1.8499 \ {\rm GeV} ,
\eqno(4.7)
$$
so that we get
$$
\frac{m_{0u}}{m_{0e}} =3.121 .
\eqno(4.8)
$$
This suggests 
$$
\frac{m_{0u}}{m_{0e}} = 3,   
\eqno(4.9)
$$
in contrast to Eq.(4.6). 
We consider that such the  
factor 3 originates in  a slight difference between 
flavon VEV scales in the lepton sector and the 
quark sector.   
In the next subsection, we will discuss such an 
additional VEV scale difference.

\vspace{2mm}

{\bf 4.2  Slight deviation of the scale $\|\Phi_{lepton}\|$ 
from $\|\Phi_{quark}\|$}  

So far, we have not mentioned the origin of the  parameter value of 
$\rho$ defined by Eq.(3.7) and the factor 3 in Eq.(4.9).
In order to understand those numerical factors, 
in this subsection. we define additional scale 
deviation factors of $\Phi_\ell$ and $\hat{S}_\ell$ 
($\ell=e, \nu$ and $q=u, d$) from  
$\Phi_q$ and $\hat{S}_q$, respectively. 
However, since the deviations are very small
(smaller than $O(1)$), arguments in this subsection
will not give an essential influence on our main
purpose which is to estimate the scales of 
U(3)$\times$U(3)$'$.  

We consider the following VEV deviation factors, $\eta_\Phi$ and
$\eta_S$, of the lepton sector
from the quark sector: 
$$
\|(\Phi_\ell)_\circ^{\ \bullet}\|= \eta_\Phi \Lambda_2, 
\ \ \ \ \|(\hat{S}_\ell)_\bullet^{\ \bullet}\|= 
\eta_S \Lambda_1,
\eqno(4.10)
$$
which is slightly different from  $\|(\Phi_q)_\circ^{\ \bullet}\|= \Lambda_2$  
and $\|(\hat{S}_q)_\bullet^{\ \bullet}\|= \Lambda_1$. 
Here, we have consider that factors $\eta_\Phi$ and
$\eta_S$ are orders of one in contrast to factors 
$\xi_\nu$ and $\xi_{S\nu}$ with orders of $10^{-3}$ as shown later. 

The modification (4.10) gives 
$$
\|(\hat{M}_\ell)_\circ^{\ \circ}\| = (\eta_\Phi)^2 (\eta_S)^{-1} 
\|(\hat{M}_q)_\circ^{\ \circ}\| ,
\eqno(4.11)
$$
so that (4.9) demands 
$$
\frac{(\eta_\Phi)^2}{\eta_S} = \frac{1}{3}, \ \ \ i.e. 
\ \ \ 3 (\eta_\Phi)^2 = \eta_S .
\eqno(4.12)
$$
On the other hand, the VEV relation (2.11) does not hold 
unless $\|(\Phi_\ell)_\circ^{\ \bullet}\|\cdot
\|(\hat{S}_\ell)_\bullet^{\ \bullet}\| = \Lambda_2 \Lambda_1$. 
Therefore, we put additional relation
$$
\eta_\Phi \eta_S = 1,
\eqno(4.13)
$$
so that we obtain
$$
3 (\eta_\Phi)^3 = 1 \ \ \Rightarrow \ \ 
\eta_\Phi = \frac{1}{\eta_S} = \frac{1}{3^{1/3}} .
\eqno(4.14)
$$
As a result, we can give the parameter value $\rho$ defined 
in Eq.(3.7) as $3^{1/3} = 1.44$ as seen in Eq.(5.30).

\vspace{5mm}

{\large\bf 5 \ Flavons in the neutrino sector} 

We discuss Majorana mass matrix, $Y_R$,  
of the right-handed neutrinos $\nu_R$.
Note that the mass matrix $\bar{Y_R}$ is 
$({\bf 6}^*, {\bf 1})$ of U(3)$\times$U(3)$'$, i.e.
$(\bar{Y}_R)^{\circ\circ}$. 

In the present neutrino mass matrix model, 
the Dirac neutrino mass matrix $\hat{M}_\nu$ is 
given by Eq.(2.5), i.e.
$$
(\hat{M}_\nu)_\circ^{\ \circ} = \frac{\langle H_u\rangle}{\Lambda_H}
(\Phi_\nu)_\circ^{\ \bullet} 
(\hat{S}_\nu^{-1})_\bullet^{\ \bullet} (\bar{\Phi}_\nu)_\bullet^{\ \circ} .
\eqno(5.1)
$$
In this section, since we pay attention to VEV scales of flavons, 
it is important whether those flavons belong  to U(3) or 
U(3)$'$.  
From Eq.(5.1), we find that the scale of $(\hat{M}_\nu)_\circ^{\ \circ}$,
 $\|(\hat{M}_\nu)_\circ^{\ \circ}\|$, is given by
$$
|| \hat{M}_\nu || \equiv \xi_M || \hat{M}_e || =
\xi_M \frac{\langle H_u \rangle }{ \Lambda_H}
\frac{(\Lambda_2)^2}{\Lambda_1} .
\eqno(5.2)
$$
Since $\|\hat{M}_e\|$ is given by the order of 
$(\langle H_d\rangle/\Lambda_H)(\Lambda_2)^2/\Lambda_1$ and
we consider an additional seesaw (1.9), 
a  ratio of the scales $\|M^{Maj}_\nu\|/\|\hat{M}_e\|$ is given by 
$$
R_{\nu/e} \equiv \frac{\|M_\nu^{Maj}\|}{\|\hat{M}_e\|} =
(\xi_M)^2 \frac{v_{H}}{\Lambda_H} 
\frac{(\Lambda_2)^2}{\Lambda_1} 
\frac{1}{\|\bar{Y}_R \|},
\eqno(5.3)
$$
where $v_{Hu} =v_{Hd} \equiv v_H$ is defined in (2.6).  
In order to estimate the ratio (5.3), we have to build a model 
for $\bar{Y}_R$.

\vspace{2mm}

{\bf 5.1 \ VEV structure of Majorana mass matrix $\bar{Y}_R$ }

For convenience, we denote the form $\langle \bar{Y}_R \rangle$ 
by the following terms
$$
\mu_R \langle \bar{Y}_R \rangle^{\circ\circ}= \mu_R \left[
(\bar{Y}_R^{1st})^{\circ\circ} + (\bar{Y}_R^{2nd})^{\circ\circ} 
\right] ,
\eqno(5.4)
$$
with $\|(\bar{Y}_R^{1st})\| \gg \|(\bar{Y}_R^{2nd})\|$.   
In Eq.(5.4) , we denote the first and second terms 
in $\bar{Y}_R$ as $(\bar{Y}_R^{1st})$ and $(\bar{Y}_R^{2nd})$, 
but it  does not  mean that they are two new flavons. 
Considering that the mass hierarchy in the neutrino sector
is mild compared with other sectors $f=u, d, e$, we 
assume that the first term of $M_\nu^{Maj}$, 
$\hat{Y}_\nu (\bar{Y}^{1st}_R)^{-1}\bar{Y}_\nu$ 
takes a form of diagonal matrix, and the observed mass differences  
and the PMNS mixing come form in the second term 
$\bar{Y}_R^{2nd}$.
Since $\langle \hat{Y}_\nu\rangle \propto Z^2$ 
($Z$ is defined in (4.4)), we take a form 
$\bar{Y}_R^{1st} \propto Z^2$.

Since we want that the $R$ charge of $\bar{Y}_R$ is as 
smaller as possible in order to avoid appearance of 
many combinations with the same $R$ charge, 
we take a form of $(\bar{Y}_R^{1st})^{\circ\circ}$ 
with  $R =13/3$, 
$(\bar{\Phi}_e^T) ^\circ_{\ \bullet} \bar{E}^{\bullet\bullet}
(\bar{\Phi}_e)_{\bullet}^{\ \circ} (\Phi_u)_\circ^{\ \bullet} 
(\bar{\Phi}_u)_\bullet^{\ \circ}$. i.e. 
$$
\mu_R (\bar{Y}_R^{1st})^{\circ\circ}  =  \frac{1}{2}
\frac{1}{(\Lambda_1)^2} \left\{ 
(\bar{\Phi}_e^T)^\circ_{\ \bullet} \bar{E}^{\bullet\bullet} 
(\hat{Y}_{eu})_\bullet^{\ \bullet}
(\bar{\Phi}_u)_\bullet^{\ \circ} + (transposed) \right\} .
\eqno(5.5)
$$
Here, in order to adjust the scale of $\bar{Y}_R^{1st}$ 
compared with $\bar{Y}_R^{2nd}$, we introduce  
a new flavon $\hat{Y}_{eu}$ defined by 
$$
\mu_{eu} (\hat{Y}_{eu})_\bullet^{\ \bullet} = 
(\bar{\Phi}_e)_\bullet^{\ \circ} (\Phi_u)_\circ^{\ \bullet} .
\eqno(5.6)
$$
Then, we obtain
$$
\mu_R \|(\bar{Y}_R^{1st})^{\circ\circ} \| = (\eta_\Phi)^2 
 \frac{\Lambda_1}{\mu_{eu}} \frac{(\Lambda_2)^4}{(\Lambda_1)^2} .
\eqno(5.7)
$$ 

Next, we discuss a term $\bar{Y}_R^{2nd}$.
We suppose that the deviation term $\bar{Y}_R^{2nd}$ gives 
Pontcorvo-Maki-Nakagawa-Sakata (PMNS) \cite{PMNS}  mixing 
and neutrino mass ratios.  
Since we want to inherit the form of $\langle \bar{Y}_R \rangle$ 
from the previous model, 
in which the form of $Y_R$ has included a term proportional 
to up-quark mass matrix $\hat{M}_u$ \cite{Yu_in_Mnu_08}.   
However, in the present model, there is no Yukawaon $\hat{Y}_u$.
Therefore, we introduce a new flavon whose VEV is proportional
to $\langle \hat{S}_u^{-1} \rangle$ with use of a superpotential given by 
$$
W_{S'_u} = \left\{ \bar{E}^{\circ\bullet} 
(\hat{E})_\bullet^{\ \bullet}  + (\bar{S}'_u)^{\circ\bullet}
 (\hat{S}_u)_\bullet^{\ \bullet} 
\right\} (\Theta'_{Su})_{\bullet\circ} ,
\eqno(5.8)
$$
where, for simplicity, we have drop the coefficients $\lambda_1$
and $\lambda_2$ although we suppose 
$\lambda_1 \simeq \lambda_2 \simeq 1$. 
Here, since $R(\hat{E}_\bullet^{\ \bullet}) = 10/5$ and 
$R( E_{\circ\bullet})=1/3$, 
the flavon $\hat{\Theta}'_{Su}$ has zero VEV value and 
$R$ charge $R=0$.  
The flavon $(\bar{S}'_u)^{\circ\bullet}$ with 
$R=5/3$ can obtain the VEV form proportional to
$\langle \hat{S}_u^{-1}\rangle$.  
Thus, we take a small deviation term $\bar{Y}_R^{2nd}$ as follows:
$$
\mu_R (\bar{Y}_R^{2nd})^{\circ\circ} = \frac{1}{2} 
\frac{1}{(\Lambda_1)^2} \left\{ 
(\bar{\Phi}_d)^\circ_{\ \bullet} (\bar{\Phi}_u^T)^\bullet_{\ \circ}
(\bar{S}'_u)^{\circ\bullet} (\bar{\Phi}_u)_\bullet^{\ \circ} 
+(transposed) 
\right\}  .
\eqno(5.9)
$$
Then, we estimate of a scale of (5.9) as
$$
\mu_R \|(\bar{Y}_R^{2nd})\| = \frac{(\Lambda_2)^4}{(\Lambda_1)^2} .
\eqno(5.10)
$$
Therefore, we obtain the ratio
$$
R_{2/1} \equiv \frac{\|\bar{Y}_R^{2nd}\|}{\|\bar{Y}_R^{1st}\|} 
= \frac{1}{(\eta_\Phi)^2} \frac{\mu_{eu}}{\Lambda_1} . 
\eqno(5.11)
$$

\vspace{2mm} 

{\bf 5.3 Parameter $\xi_R$ }

Parameter fitting for neutrino data such as neutrino masses 
and PMNS lepton mixing matrix is done under the following
dimensionless re-expression:
$$
\tilde{M}_\nu^{Maj} = (\xi_M)^4 Z^2 \tilde{Y}_R^{-1} 
Z^2 ,
\eqno(5.12)
$$
where
$$
\tilde{Y}_R = Z^4 + \xi_R 
Z^2 {P} ({\bf 1} + b_u X_3 )^{-1} 
{P}^\dagger Z, 
\eqno(5.13)
$$
$$
\begin{array}{l}
Z = {\rm diag} (z_1, z_2, z_3)  , \\
{P} = {\rm diag} ( e^{i \tilde{\phi}_1},  e^{i \tilde{\phi}_2}, 1) .
\end{array}
\eqno(5.14)
$$
($Z$ and ${P}$ do not mean new flavons. 
Those are noting but dimensionless $3\times 3$ matries.)   
The parameter $\xi_R$ corresponds to the ratio $R_{2/1}$ defined 
in (5.11).
Since the parameter values $b_u$ and 
$(\tilde{\phi}_1,  \tilde{\phi}_2)$ 
have already determined from the CKM fitting as shown in (2.25) 
and (2.26), only a free parameter in the neutrino sector is 
$\xi_R$ in (5.13). 
(The parameter fitting is practically the same 
as one in the previous model  \cite{K-N_PRD15R, K-N_MPLA16}  
However, note that the present parameter $\xi_R$ corresponds 
to $2/\xi_R$ in the previous model.)  
The best fitting value of $\xi_R$ is \cite{K-N_PRD15R, K-N_MPLA16}
$$
\xi_R = 0.9806 \times 10^{-3} .
\eqno(5.15)
$$
The detail fitting is reviewed in Appendix A. 

\vspace{2mm} 

{\bf 5.4 Scales of $\mu$ parameters }

We have four flavon mass parameters $\mu_\nu$,
$\mu_{S\nu}$, $\mu_{eu}$ and $\mu_R$  defined by (2.18), 
(2.22), (5.6) and (5.5), respectively.
So far, we have considered three VEV scales $\Lambda_1$, 
$\Lambda_2$ and $\Lambda_3$ as shown in (1.8) and also 
in Table 1.
Therefore, let us  put the following selection rules for 
$\mu$ parameters defined by
$$
\mu_A \|A\| = \|G\|, 
\eqno(5.16)
$$
where $G$ is a combination of some flavons including 
factor $1/(\Lambda_1)^n$ :
(i) We assume that our parameters $\mu_A$ such as $\mu_\nu$, $\mu_{S\nu}$, 
$\mu_{eu}$ and $\mu_R$ are given by some of those 
three scales $\Lambda_1$, $\Lambda_2$ and $\Lambda_3$. 
(ii) If our choice $\mu_A= \Lambda_a$
($a=1,2,3$) , as a result, gives a VEV value of $\|A\|$ which is 
larger than the maximally allowed scale of $\|A\|$, e.g. 
$\Lambda_1$ for $\|(\hat{Y}_{eu})_\bullet^{\ \bullet}\|$, 
$\Lambda_2$ for $\|({\Phi}_\nu)_\circ^{\ \bullet}\|$, 
and $\Lambda_3$ for $\|(\bar{Y}_R)^{\circ \circ}\|$, 
then, the choice  is ruled out. 
(iii) If the choice $\mu_A =\Lambda_a$ gives a 
VEV value $\|A\|$ which is two rank lower compared 
with $\|A\|_{max}$, the choice is also ruled out.

For example, let us see the case (5.6):
$$
\mu_{eu} \|(\hat{Y}_{eu})_\bullet^{\ \bullet}\| = 
\eta_\Phi (\Lambda_2)^2 = \eta_\Phi  \rho\Lambda_1 \Lambda_3 = \eta_\phi (\Lambda_2)^2 . 
\eqno(5.17)
$$
If we tale $\mu_{eu}=\Lambda_3$, we obtain 
$\| \hat{Y}_{eu}\| = \eta_\Phi \rho \Lambda_1$ from Eq.(5.17).
Since $\|(\hat{Y}_{eu})_\bullet^{\ \bullet}\|\leq 
\Lambda_1$, the case $\mu_{eu}=\Lambda_3$ is not rulded 
out by the rule (ii).
However, the case give 
$R_{2/1} = (\eta_{\Phi})^{-2} \sim O(1)$ from (5.11). 
This contradicts with the parameter fitting result (5.15).
Therefore, we rule out this case $\mu_{eu}=\Lambda_3$. 
On the other hand, if we take $\mu_{eu} =\Lambda_1$, 
we obtain $\|(\hat{Y}_{eu})_\bullet^{\ \bullet}\|= \rho \Lambda_3$. 
However, the value is two rank small compared with the maximal 
value $\Lambda_1$, so that we rule out the case $\mu_{eu}=\Lambda_1$
by the rule (iii).  
As a result, we choose $\mu_{eu} = \Lambda_2$,
and then we obtain
$$
\|(\hat{Y}_{eu})_\bullet^{\ \bullet}\|= \eta_\Phi \Lambda_2 .
\eqno(5.18)
$$
Then, from the relation (5.11), we obtain
$$
\xi_R \equiv R_{2/1} =\frac{1}{(\eta_\Phi)^2} 
\frac{\Lambda_2}{\Lambda_1} .
\eqno(5.19)
$$

Similarly, from the relation (2.18), we have
$$
\mu_\nu \|(\Phi_\nu)_\circ^{\ \bullet}\| = 
\eta_\Phi \frac{(\Lambda_2)^2}{\Lambda_1} \Lambda_2 =
 \eta_\Phi \rho \Lambda_3 \Lambda_2 =
\rho \Lambda_3 \|(\Phi_e)_\circ^{\ \bullet}\| .
\eqno(5.20)
$$
If we take take $\mu_\nu =\Lambda_3$, we obtain 
$\| (\Phi_\nu)_\circ^{\ \bullet}\| = \rho  \|(\Phi_e)_\circ^{\ \bullet}\|$, i.e. 
$\xi_\nu = \rho >1$. 
The result is not our desired one, because our aim 
is to understand tiny neutrino masses by $\xi_\nu \ll 1$.  
Therefore, we choose $\mu_\nu =\Lambda_2$, and we get
$$
\| (\Phi_\nu)_\circ^{\ \bullet}\|= \eta_\Phi \rho \Lambda_3 
= \rho \frac{\Lambda_3}{\Lambda_2} \eta_\Phi \Lambda_2 
= \frac{\Lambda_2}{\Lambda_1} \| (\hat{S}_e)_\bullet^{\ \bullet}\|,
\eqno(5.21)
$$
so that we have
$$
\xi_\nu = \frac{\Lambda_2}{\Lambda_1} =(\eta_\Phi)^2 \xi_R .
\eqno(5.22)
$$
from (5.11).

Similarly, when we choose $\mu_{S\nu}= \Lambda_2$ in Eq.(2.23), 
from 
$$
\mu_{S\nu} (\hat{S}_\nu)_\bullet^{\ \bullet} = 
\eta_S \Lambda_1 \frac{(\Lambda_2)^2}{\Lambda_1} 
= \eta_S \rho \Lambda_1 \Lambda_3 ,
\eqno(5.23)
$$
we obtain
$$
\|(\hat{S}_\nu)_\bullet^{\ \bullet}\| = \eta_S 
\frac{\Lambda_2}{\Lambda_1}\Lambda_1 =
\frac{\Lambda_2}{\Lambda_1} \|\hat{S}_e\| ,
\eqno(5.24)
$$
so that we have
$$
\xi_{S\nu} = \frac{\Lambda_2}{\Lambda_1} = 
(\eta_\Phi)^2  \xi_R = \xi_\nu ,
\eqno(5.25)
$$
from Eq.(5.22). 
Therefore, from the relation (2.24), we obtain
$$
\xi_M = (\xi_\nu)^2 (\xi_{S\nu})^{-1} = \xi_\nu =
(\eta_\phi)^2 \xi_R .
\eqno(5.26)
$$

Finally, we discuss a scale of $\mu_R$.  
From (5.7), by regarding $Y_R$ as $Y_R \simeq Y_R^{1st}$,  
we can write 
$$
\mu_R \|(\bar{Y}_R)^{\circ\circ}\| = 
\frac{1} {(\Lambda_1)^2} (\eta_\Phi)^2 
\Lambda_2 \Lambda_1 \Lambda_2 \Lambda_2 = 
\rho (\eta_\Phi)^2 \Lambda_2 \Lambda_3 .
\eqno(5.27)
$$
If we take $\mu_R = \Lambda_2$, we obtain 
$$
\|(\bar{Y}_R)^{\circ\circ}\| = 
\rho  (\eta_\Phi)^2 \Lambda_3 .
\eqno(5.28)
$$
Since $\|(\bar{Y}_R)^{\circ\circ}\|$ cannot have 
a larger scale than $\Lambda_3$, we have a constraint 
$$
\rho  (\eta_\Phi)^2 \leq 1, \ \ i.e. \ \ 
\rho \leq (\eta_\Phi)^{-2} = 3^{2/3} .
\eqno(5.29)
$$
Comparing the fitting value $\rho=1.37$ with the value
$(\eta_\Phi)^{-1} = 3^{1/3} = 1.44$ in (4.14), it is likely that 
the value of $\rho$ is given in unit of $3^{1/3}$, 
so that we regard the value 
of $\rho$ as  
$$
\rho = \frac{1}{\eta_\Phi} = 3^{1/3} .
\eqno(5.30)
$$
Then, this choice satisfies the condition 
$\|Y_R\| \leq \Lambda_3$ in (5.28).

\vspace{5mm}

{\large\bf 6 \ Estimate of flavon VEV scales}

Similarly to Eq.(4.1), from the diagonalization 
of (5.12) with the parameter values (2.25), (2.26) 
and (5.15), we obtain 
$$
(m_{\nu 1}, m_{\nu 2}, m_{\nu 3}) = 
(0.30892, 0.31689, 0.51026)\, m_{0\nu} .
\eqno(6.1)
$$
Here, we have defined $m_{0R}$ by
$$
m_{0\nu} = (\xi_M)^2 \frac{(m_{0e})^2}{m_{0R}} . 
\eqno(6.2)
$$
From the observed values \cite{PDG14} 
$\Delta m_{32}^2=0.000244$ eV$^2$ and 
$\Delta m_{21}^2/\Delta m_{32}^2 = 0.0309$, 
we estimate $m_{\nu 3} = 0.063$ eV. Thus we obtain
$$
m_{0\nu} = 1.235 \times 10^{-13} \ {\rm TeV} . 
\eqno(6.3)
$$
Then, from Eq.(6.2) with (6.3) and (4.7),  we obtain , 
$$
m_{0 R} = (\xi_M)^2 \frac{(m_{0e})^2}{m_{0\nu}}
= (\xi_M)^2 \times 2.7710 \times 10^7 \ {\rm TeV} .
\eqno(6.4)
$$
From Eq.(5.26) and (4.14), we estimate 
$$
m_{0R} = (\eta_\Phi)^4 (\xi_R)^2  \times 2.7710 \times 10^7 \ {\rm TeV}
= 6.158 \ {\rm TeV}.
\eqno(6.5)
$$ 
Therefore, from (5.28) with (5.30), i.e.  
$\|Y_R\| = \eta_\Phi \Lambda_3$,   
we obtain 
$$
\Lambda_3 = (\eta_\Phi)^{-1} m_{0R} = 8.883 \ {\rm TeV} .
\eqno(6.6)
$$ 

In conclusion, we obtain
$$
\begin{array}{l}
\Lambda_3 =8.883 \ {\rm TeV} , \\
\Lambda_2 =\rho \Lambda_3/\xi_R = 1.306 \times 10^4 \ {\rm TeV} , \\ 
\Lambda_1 = \Lambda_2/\xi_R =1.332 \times 10^7 \ {\rm TeV}.
\end{array}
\eqno(6.7)
$$

Finally, we estimate the value of $\Lambda_H$ defined in (2.5).
From (2.5), we use a relation
$$
m_{0e} = \| \Psi_e (\hat{S}_e)^{-1} \bar{\Phi}_e \|
=  \frac{v_{Hd}}{\Lambda_H} (\eta_\Phi)^2 (\eta_S)^{-1}
\frac{(\Lambda_2)^2}{\Lambda_1} = 
\frac{v_H}{\Lambda_H} (\eta_\Phi)^2 (\eta_S)^{-1} \rho \Lambda_3,
\eqno(6.8)
$$
where $m_{0e} = 1.8499 \times 10^{-3}$ TeV, 
$v_H = 173.9\times 10^{-3}$ TeV, and  
$(\eta_\Phi)^2 (\eta_S)^{-1} \rho = 3^{-2/3}$, 
so that we obtain
$$
\Lambda_H = 401.4 \ {\rm TeV} .
\eqno(6.9)
$$
The result (6.9) gives 
$ (v_H)/\Lambda_H = 0.4332 \times 10^{-3}$.


\vspace{5mm}

{\large\bf 7 \ Concluding remarks} 

In conclusion, we have investigated a unified quark and lepton 
mass matrix model on the basis of U(3)$\times$U(3)$'$ family 
symmetry. 
We have inherited a basic aim of a series of the so-called
Yukawaon models \cite{Yukawaons}. 
However, in the present model, we have not assumed existence 
of Yukawaons. 
Instead, we have introduced triplet fermions $F_\alpha$ of
U(3)$'$ in addition to quarks and leptons $f_i$ which are 
triplets of U(3), and we assumed a seesaw-like mechanism in (2.1).

The U(3)$'$ symmetry is broken into S$_3$ at a scale 
$\mu=\Lambda_1$, and the U(3) symmetry is broken at 
$\mu=\Lambda_2$ by VEVs of flavons 
$\langle (\Phi_f)_i^{\ \alpha}\rangle$, 
which are $({\bf 3}, {\bf 3}^*)$ of U(3)$\times$U(3)$'$. 
However, in this paper, we do not ask the origin of the 
VEV form $\langle \Phi_f \rangle \propto  {\rm diag}
(\sqrt{m_e}, \sqrt{m_\mu}, \sqrt{m_\tau})$, and do not 
discuss what mechanism leads to such a VEV form.
We consider that it is too early to investigate the origin 
of the values $m_{ei}$. 
It is our future task.

The purpose of the present paper is not to give parameter 
fitting for quark and lepton masses and mixing because 
the VEV structures are the same 
as previous model, although  the present model is completely different from the previous one. 
In the present model, we have only six free family-number-independent parameters 
for quark and lepton mass ratios and mixing as well as 
in the previous model.  
We have used the values $b_u=-1.011$, 
$b_d= -3.522\, e^{i\, 17.7^\circ}$, 
$(\tilde{\phi}_1, \tilde{\phi}_2)$  
and $\xi_R = 0.4903 \times 10^{-3}$,  
whose values have been quoted from the previous paper
\cite{K-N_PRD15R}. 
  
In this paper, we have investigated 
the symmetry-breaking scales $\Lambda_1$ 
and $\Lambda_2$ from phenomenological study 
in the neutrino data. 
Our essential hypothesis is that all flavon VEVs 
(and also $\mu$ parameters) take one of three scales 
$\Lambda_1$, $\Lambda_2$ and $\Lambda_3$ except for 
$H_{u/d}$ and $\Lambda_H$. 
First, we have fixed the ratio 
$\Lambda_2/\Lambda_1 \sim \Lambda_3/\Lambda_2 
\sim 10^{-3}$  from the $\phi_i$-$m_{ei}$ relation 
with observed data as discussed in Sec.3. 
Here, we would like to emphasize that the relations (3.3) 
with the numerical fitting values (3.5) are essential 
for our result $\Lambda_2/\Lambda_1 \sim \Lambda_3/\Lambda_2$, 
and the relations (3.3) are completely different from that 
in the previous model \cite{K-N_PRD15R, K-N_MPLA16}.  
Then, we have obtain the value of  
$\Lambda_1 \sim 4\times 10^7$ TeV,  
$\Lambda_2 \sim 2\times 10^4$ TeV, and 
$\Lambda_3 \sim 10^1$ TeV.
Here, the value $\xi_R \sim 10^{-3}$ has been obtained for
the parameter  $\xi_R$ which is only the 
free parameter in the neutrino sector after 
remaining free parameters have been fixed by 
fitting the observed quark mass ratios and CKM mixing. 
The scale $\Lambda_3$ has been fixed from 
the scale of the Majorana mass matrix 
$\langle (\bar{Y}_R)^{\circ\circ} \rangle$. 

Note that our estimate of $\Lambda_1$, 
$\Lambda_2$ and $\Lambda_3$ is highly dependent 
on the VEV structure of the flavon $\bar{Y}_R$, 
which corresponds to the Majonara mass matrix of 
right-handed neutrinos $\nu_R$.  
We have inherited the VEV structure from the previous model, 
which has given excellent description 
of neutrino masses and mixing with only one parameter
$\xi_R$ phenomenologically. 
The structure is dependent on the $R$ charge assignment. 
The present $R$ charge assignment $(r_u, r_e, r_d, r_\nu)=(1/3, 1, 6/3,
10/3)$ is one of a few possible cases which do not cause  
theoretical trouble, although it has no theoretical
ground.  
Investigation of the logical necessity of this assignment 
is a future task to us.

In Sec.4, we have discussed somewhat trifling 
parameterization with an additional ansatz. 
The orders of three VEV scales can roughly be 
obtained without these parameters in Sec.4.   
However, owing to this additional parametrization, 
we can understand $m_{0u}/m_{0e}=3$ and 
$\rho \equiv c_2/c_1 \simeq  3^{1/3}$ consistently.  

Although the present model still leaves some of tasks 
in future, we consider that the outline of the model is 
worthy to notice.

\vspace{5mm}

{\large\bf Acknowledgments}

The authors thank T.~Yamashita and Y.~Sumino for 
helpful discussions on the effective superpotential and 
helpful comments. 
They also thank T.~Asaka and K.~Kaneta for valuable comments
on a light right-handed Majorana neutrino mass and 
$\nu_L$-$\nu_R$ mixing. 
The work was supported by JSPS KAKENHI Grant
number JP16K05325 (Y.K.).

\vspace{5mm}

{\large\bf Appendix A: Parameter fitting for masses and mixings} 

\begin{table}
\caption{Predicted values vs. observed values in quark sector. 
Input values are given by Eq.(2.25) and (2.26).
$r^q_{12}$ and $r^q_{23}$ are defined by 
$r^{u}_{12}= \sqrt{m_{u}/m_{c}}$, 
$r^{u}_{23}= \sqrt{m_{c}/m_{t}}$, 
$r^{d}_{12}= m_{d}/m_{s}$
and $r^{d}_{23}= m_{s}/m_{b}$.  
The predicted values are same as those in the previous paper\cite{K-N_PRD15R}. }

\vspace*{2mm}
\begin{tabular}{|c|ccccccccc|} \hline
  & $|V_{us}|$ & $|V_{cb}|$ & $|V_{ub}|$ & $|V_{td}|$ & 
$\delta^q_{CP}(^\circ)$ &  $r^u_{12}$ & $r^u_{23}$ & $r^d_{12}$ & $r^d_{23}$ 
 \\ \hline 
Predicted &$0.2257$ & $0.03996$ & $0.00370$ & $0.00917$ & $81.0$ & 
$0.061$ & $0.060$ & $0.049$ & $0.027$ 
 \\
Observed & $0.22536$ & $0.0414$ &  $0.00355$  & $0.00886$  & $69.4$ &
$0.045$ & $0.060$ & $0.053$  & $0.019$  
  \\
    &  $ \pm 0.00061$ &  $ \pm 0.0012$& $ \pm 0.00015$ & 
 ${}^{+0.00033}_{-0.00032}$ & $\pm 3.4$ &
${}^{+0.013}_{-0.010}$ & $ \pm 0.005$ & $^{+0.005}_{-0.003}$ &
${}^{+0.006}_{-0.006}$ 
 \\ \hline
\end{tabular}
\end{table}

Since the purpose of the present paper is not to give parameter 
fitting of masses and mixings of quarks and leptons, so far, 
we have not give the details of the parameter fittings.
The present is drastically changed from the previous model 
\cite{K-N_PRD15R, K-N_MPLA16}.
Nevertheless, the main part of the VEV relations are substantially 
the same as the previous those. 
Therefore, we have quoted the fitting values $(bu, b_d)$ in Eq.(2.25),
$(\tilde{\phi}_1, \tilde{\phi}_2)$ in Eq.(2.26) and $\xi_R$ in Eq.(5.15) 
from the previous paper \cite{K-N_PRD15R, K-N_MPLA16}.  
However, here, in order to help for reader's understanding, 
we would like to give  a brief review of the parameter fitting.

First, note that the parameter fitting in Ref.\cite{K-N_PRD15R, K-N_MPLA16}
has been done at $\mu = m_Z$.  (For the charged lepton 
mass values, too, we have used the values at $\mu=m_Z$, 
not pole mass values.)  

In our model, the parameters $(b_u, b_d)$ and 
$(\tilde{\phi}_1, \tilde{\phi}_2)$ 
are fixed by quark masses and CKM quark mixing fitting. 
We have chosen the parameter values
as
$$
b_u = -1.011 , \ \ \ \ b_d = -3.522 \, e^{i\ 17.7^\circ},            
\eqno(A.1)
$$
and
$$
(\tilde{\phi}_1, \tilde{\phi}_2 )= ( -176.05^\circ,\ -167.91^\circ). 
\eqno(A.2)
$$
Then, we can obtain reasonable quark mass ratios and CKM mixing
as shown in Table.2.

Parameter fitting for neutrino data is done under the following
dimensionless re-expression:
$$
\tilde{M}_\nu^{Maj} = (\xi_M)^4 Z^2 \tilde{Y}_R^{-1} 
Z^2 ,
\eqno(A.3)
$$
where
$$
\tilde{Y}_R = Z^4 + \xi_R 
Z^2 {P} ({\bf 1} + b_u X_3 )^{-1} 
{P}^\dagger Z, 
\eqno(A.4)
$$
$$
\begin{array}{l}
Z = {\rm diag} (z_1, z_2, z_3)  , \\
{P} = {\rm diag} ( e^{i \tilde{\phi}_1},  e^{i \tilde{\phi}_2}, 1) .
\end{array}
\eqno(A.5)
$$
($Z$ and ${P}$ do not mean new flavons. 
Those are noting but dimensionless $3\times 3$ matries.)   

Since the parameter values $b_u$ and 
$(\tilde{\phi}_1,  \tilde{\phi}_2)$ 
have already determined from the CKM fitting as shown in (A.1) 
and (A.2), only a free parameter in the neutrino sector is 
$\xi_R$ in (A.4) as discussed in subsection 5.3. 
(The parameter fitting is practically the same 
as one in the previous model  \cite{K-N_PRD15}.  
However, note that the present parameter $\xi_R$ corresponds 
to $2/\xi_R$ in the previous model.)

\begin{figure}[ht]
\begin{picture}(200,200)(0,0)

  \includegraphics[height=.33\textheight]{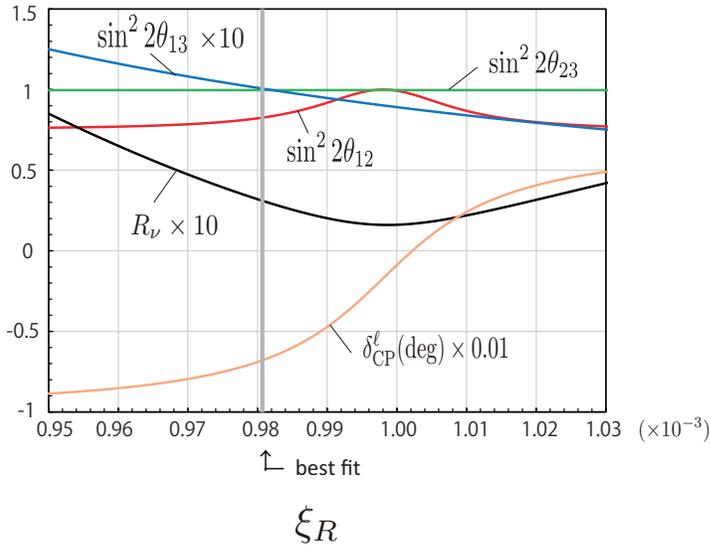}

\end{picture}  
  \caption{$\xi_R$ dependence of the lepton mixing parameters 
$\sin^2 2\theta_{12}$, 
$\sin^2 2\theta_{23}$, $\sin^2 2\theta_{13}$, and the neutrino 
mass squared difference ratio 
$R_{\nu} \equiv {\Delta m_{21}^2}/{\Delta m_{32}^2}$. 
For the input values of  $b_u$ and $(\phi_1, \phi_2)$, 
see the text.
} 
 \label{fig1}
\end{figure}

Therefore, as shown in the previous paper, 
the PMNS mixing parameters $\sin^2 2\theta_{12}$, $\sin^2 2\theta_{23}$, 
$\sin^2 2\theta_{13}$, 
$CP$ violating Dirac phase parameter $\delta_{\rm CP}^\ell$,  and 
the neutrino mass squared difference ratio 
$R_{\nu} \equiv {\Delta m_{21}^2}/{\Delta m_{32}^2}$  are 
turned out to be functions 
of the remaining only one  parameter $\xi_R$. 
In Fig.~1, we draw the curves as  functions of our new  $\xi_R$
with taking  $b_u=-1.011$, and  $b_d=-3.3522$, $\beta_d=17.7^\circ$, and 
$(\phi_1, \phi_2)=(-176.05^\circ, -167.91^\circ )$. 
From Fig.1, 
it turns out that 
the best fitting value of $\xi_R$ in the present paper is 
$$
 \xi_R = 0.9806 \times 10^{-3},  \quad  \quad (\,\,\frac{2}{\xi_R}=2039.6\,\, ), 
\eqno(A.6)
$$
which gives the predictions 
$$
\sin^2 2\theta_{12} = 0.8254, \ \ 
\sin^2 2\theta_{23} =0.9967, \ \ \sin^2 2\theta_{13} = 0.1007, \ \ 
 \delta_{\rm CP}^\ell =-68.1^\circ,  \ \  R_{\nu} = 0.03118,
\eqno(A.7)
$$
as shown in Table~3. 
These predictions are in good agreement with
the observed values \cite{PDG14}.

The neutrino masses are predicted as 
$m_{\nu 1} \simeq 0.038$ eV, $m_{\nu 2} \simeq 0.039$ eV, and
$m_{\nu 3} \simeq 0.063$ eV,
by using the input value \cite{PDG14}
$\Delta m^2_{32}\simeq 0.00244$ eV$^2$.
We have also predicted the effective Majorana neutrino mass \cite{Doi1981} 
$\langle m \rangle$ 
in the neutrinoless double beta decay as
$$
\langle m \rangle =\left|m_{\nu 1} (U_{e1})^2 +m_{\nu 2} 
(U_{e2})^2 +m_{\nu 3} (U_{e3})^2\right| 
\simeq 21\  \mbox{ meV}.
\eqno(A.8)
$$

\vspace{2mm}

\begin{table}
\caption{Predicted values vs. observed values in neutrino sector. 
} 

\vspace*{2mm}
\hspace*{-17mm}
\begin{tabular}{|c|ccccccccc|} \hline
   & $\sin^2 2\theta_{12}$ & $\sin^2 2\theta_{23}$ & $\sin^2 2\theta_{13}$ & 
 $R_{\nu}\ (10^{-2})$ &  
$\delta^\ell_{CP}(^\circ)$ & $m_{\nu 1}\ ({\rm eV})$ & $m_{\nu 2}\ ({\rm eV})$ & 
$m_{\nu 3}\ ({\rm eV})$ & $\langle m \rangle \ ({\rm eV})$ \\ \hline
 Predicted & $0.8254$ & $0.9967$ & $0.1007$ &  $3.118$ & $-68.1$ &
 $0.038$ & $0.039$ & $0.063$ & $0.021$ \\
Observed & $0.846$   & $ 0.999$ & $0.093$ &   $3.09 $    &no data
  &no data  &no data  &no data  &  $<\mathrm{O}(10^{-1})$   \\ 
    & $ \pm 0.021$ & $^{+0.001}_{-0.018}$   & $\pm0.008$ & $ \pm 0.15 $  &  &
   &    &    &    \\ \hline 
\end{tabular}
\end{table}


It is interesting that the model predicts 
$\delta_{CP}^\ell = -68^\circ$, which shows 
$\delta_{CP}^\ell \simeq -\delta_{CP}^q$.
It is also worthwhile noticing that we obtain a large value
21 meV for the effective Majorana neutrino mass $\langle m\rangle$  
in spite of the normal hierarchy for the neutrino mass in our model.



\vspace{5mm}
%


\end{document}